\documentclass[12pt,english]{article}
\usepackage{babel} % DEUTSCHE SONDERZEICHEN
\usepackage[ansinew]{inputenc} % UMLAUTE
\usepackage{setspace} % ZEILENABST\HAT{\THETA_I}NDE USW
\usepackage{amsmath} % MATHE
\usepackage{amssymb} % MATHE
\usepackage{comment}
\usepackage{bbm}
\usepackage{amsfonts} % MATHE
\usepackage{titling}
\usepackage[small]{caption}
\usepackage{geometry}
\usepackage{multirow,array}
\usepackage{amsthm}
\usepackage{epic,eepic}
\usepackage{hyperref}
\usepackage{multirow}
\usepackage{bbm}
\usepackage{subcaption}
\usepackage{dcolumn}
\usepackage{adjustbox}
\usepackage{breakurl}
\usepackage{xurl}
\usepackage{rotating}
\usepackage{array}
\usepackage{natbib}

\newcommand{\ot}{\frac{1}{2}}

\usepackage{breakcites}
\usepackage{graphicx}
\usepackage[UKenglish]{datetime}
\usepackage{color}
\newtheorem{prop}{Proposition}

\selectfont

\begin{document}

\title{\textbf{Does Ideological Polarization Lead to Policy Polarization?}\thanks{
I would like to thank Nicolas Motz and the seminar audience at EPSA 2025 for valuable comments and suggestions.
Financial support of the Agencia Estatal de Investigaci\'{o}n (Spain) through grants PID2022-
141823NA-I00, MICIN/ AEI/10.13039/501100011033, and  CEX2021-001181-M as well as by Comunidad de Madrid through grant EPUC3M11 (V PRICIT)  is gratefully acknowledged.}}
\author{Philipp Denter\thanks{Universidad Carlos III de Madrid, Department of Economics, Calle de Madrid 126, 29803 Getafe, Spain. E-Mail: \href{mailto:pdenter@eco.uc3m.es}{pdenter@eco.uc3m.es}.} }

\maketitle

\begin{abstract}
I study electoral competition between two ideologically polarized parties that are both office- and policy-motivated. Parties have fixed and opposing ideological positions but compete by choosing platforms on a second, continuous policy dimension. The analysis uncovers a non-monotonic relationship between ideological and policy polarization. When ideological polarization is low, an increase leads to policy moderation; when it is high, the opposite occurs, and policies become more extreme.  Moreover, an electoral advantage shifts both parties' platforms toward the advantaged party's preferred policy, implying that the advantaged party adopts the more extreme position. More generally, both high- and low-valence candidates may adopt more extreme policies, depending on the electorate's degree of ideological polarization.
\end{abstract}

 \vspace{1cm}

\noindent \emph{JEL Codes}:
D70, 
D72,
%D72 	Political Processes: Rent-Seeking, Lobbying, Elections, Legislatures, and Voting Behavior\\
D78,
%%Positive Analysis of Policy Formulation and Implementation 

\noindent \emph{Keywords}:   electoral competition, polarization, differentiated candidates, ideology, valence

\newpage
\section{Introduction}

Many democratic societies have become significantly more polarized in recent decades. For example, \cite{AbramowitzSaunders:2008} document that ``\emph{ideological polarization has increased dramatically among the mass public in the United States as well as among political elites}.'' \cite{BoxellGentzkowShapiro:2024} show rising polarization since the 1980s in the United States, Canada, Switzerland, France, and New Zealand, while \cite{CarothersODonohue:2019} report similar patterns in Kenya, Indonesia, Turkey, and Poland. It is widely believed that such polarization undermines democratic decision making by complicating learning processes and obstructing compromise between parties. Given these concerns, it is natural to ask how rising ideological polarization affects the policies that parties put forward. Does it necessarily translate into more polarized platforms, or might it instead have a moderating effect? 

\begin{figure}[htbp]
\centering

\begin{minipage}[t]{0.48\textwidth}
  \centering
  \includegraphics[width=\textwidth]{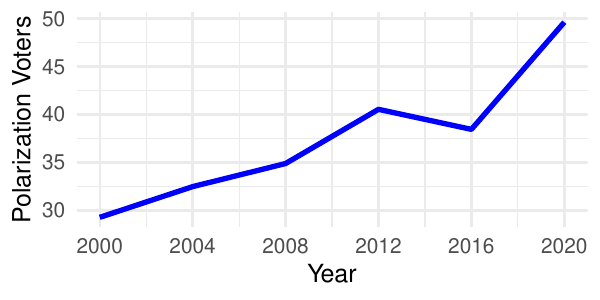}
  \caption*{(a) Affective polarization of voters.}
\end{minipage}
\hfill
%\begin{minipage}[t]{0.32\textwidth}
%  \centering
%  \includegraphics[width=\textwidth]{HOUSE_POL.pdf}
%  \caption*{(b) Ideological polarization House of Representatives.}
%\end{minipage}
%\hfill
\begin{minipage}[t]{0.48\textwidth}
  \centering
  \includegraphics[width=\textwidth]{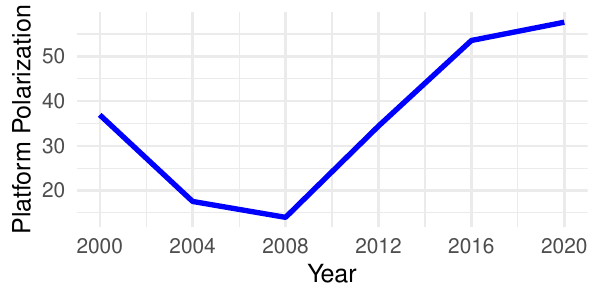}
  \caption*{(b) Platform polarization.}
\end{minipage}

\caption{Different measures of polarization in the United States since 2000.}
\label{fig:side_by_side}
\end{figure}

Empirically, the answer appears ambiguous. Figure~\ref{fig:side_by_side} displays different measures of polarization in the United States since 2000. The left panel (a) shows affective polarization of voters, while the right panel (b) displays  polarization of policy platforms.\footnote{Based on own calculations. Affective polarization is measured as the average absolute difference between respondents' feeling-thermometer evaluations of the Democratic and Republican parties in the \cite{ANES2020TS}.
%Ideological polarization in the House of Representatives is measured as the difference between average \emph{DW-NOMINATE} scores (dimension 1; \citealp{Voteview}) for Republicans and Democrats. 
Policy polarization is defined as the difference between the \emph{rile} variable of Republicans and Democrats in the Manifesto Project Database \citep{Lehmann:2024}.}

The figure shows that affective  polarization has increased almost continuously, whereas policy polarization first declined and then began rising again around 2008. This pattern suggests a non-monotonic, U-shaped relationship between ideological and policy polarization, at least in the United States.

Motivated by these observations, the paper develops a game-theoretic framework to study how ideological polarization shapes policy polarization. Two parties, both office- and policy-motivated and polarized along a non-policy ideological dimension, compete by offering policies to voters in order to secure election. Parties have fixed and opposing ideological positions, but are free to choose their platforms on a second, continuous policy dimension. The voter cares about both ideological and policy distance as well as candidate valence, and parties are uncertain about the voter's ideological position and about relative valence. By contrast, the voter's preferred position on the flexible policy dimension is assumed to be known. This last assumption is made for tractability and is not essential for the qualitative results; allowing for additional uncertainty about policy preferences would complicate the analysis without changing the underlying mechanism. Ideological polarization is captured by the weight voters attach to the fixed ideological dimension relative to the flexible policy dimension.

Proposition~\ref{prop:1} first isolates the two opposing forces generated by the two sources of electoral uncertainty. With only valence uncertainty, greater ideological polarization induces parties to moderate their policy platforms, whereas with only uncertainty about voters' ideological preferences, it induces greater policy divergence. Proposition~\ref{prop:sym_EQ} then shows that when both sources of uncertainty are present, equilibrium policy choices are non-monotonic in ideological polarization: moderation prevails when ideological polarization is low, while divergence prevails when it is high.  
Proposition~\ref{prop:dDeltedw} establishes that policy polarization is therefore U-shaped in ideological polarization, mirroring the empirical pattern in Figure~\ref{fig:side_by_side}. Proposition~\ref{prop:asum_polarization} further shows that this result is robust to small departures from symmetry and characterizes how such asymmetries shift equilibrium policy platforms. The paper thus provides a theoretical explanation for why ideological and policy polarization need not move in parallel.

A second implication of Proposition~\ref{prop:asum_polarization} is that a small electoral advantage shifts both equilibrium platforms toward the advantaged party's preferred policy. Hence, the advantaged party becomes more extreme, while the disadvantaged party becomes more moderate. More generally, in societies with low ideological polarization, \textit{high-valence} parties tend to adopt \textit{more extreme} policies than their valence-disadvantaged opponents. By contrast, in highly polarized societies, this pattern reverses: the valence-disadvantaged party may adopt the more extreme position if it has an ideological advantage (Proposition~\ref{prop:asym2}). 

This result stands in contrast to \cite{BuisseretvanWeelden:2022}, who reach the opposite conclusion. The difference arises because, unlike in \cite{BuisseretvanWeelden:2022}, parties here are both office- and policy-motivated and face also valence uncertainty. These features change the way an electoral advantage affects policy choice in a clear way. In their model, the advantaged party has stronger incentives to moderate in order to secure electoral support. Here, by contrast, an electoral advantage relaxes the electoral constraint faced by a policy-motivated party. The advantaged party can therefore move its platform closer to its own preferred policy while still maintaining a sufficiently high probability of winning. This is why, in the present model, the advantaged party is  the more extreme one.

%
%
%
%
%\cite{AbramowitzSaunders:2008}
%\cite{FiorinaAbrams:2008}
%

%
%
%
%\begin{figure}
%  \centering
%  \includegraphics[width=.6\textwidth]{IPol2.pdf}
%  \caption{Platform polarization over time}\label{fig:polpol}
%\end{figure}
%
%%
%
%Policy prefs uniform: polarization of policies increases in polarization of ideologies.
%Policy prefs normal: non monotonioc relationship. w small, means polarization of platforms decreases in w. but when w large, this changes. note: can solve in closed form for Delta with C[1] undetermined. this in theory could mean that Delta strictly decreases. but we know from the diff of the policy derivatives that when w large the derivative is pos. this allows us to bound C[1]
%

%
%
%
%-extend \cite{BuisseretvanWeelden:2022} to a setting with a continuum of policies.
%-extend \cite{BuisseretvanWeelden:2022} to a setting with policy preferences of candidates a la \cite{Wittman:1983}
%-ideological polarization implies policy polarization
%
%
%%manifesto project plus polarization data?
%%
%% AER-INSIGHTSD
%%

\paragraph{Literature:}
The paper contributes to several strands of literature in political economy. First and foremost, it adds to the literature studying the determinants of policy polarization. 
\cite{ALESINARosenthal:2000} show that when policies are determined by an executive-legislative compromise, voter behavior tends to moderate policies, which in turn creates incentives for parties to adopt more radical policy platforms.  \cite{carrillo2008information} show that imperfect information about candidate quality can induce policy divergence even among purely office-motivated candidates.
\cite{LevyRazin:2015QJPS} show that greater polarization of opinions caused by correlation neglect may lead to less polarization of policies. 
\cite{MatakosTrompounisXefteris:2015, MatakosTrompounisXefteris:2016} study how different degrees of disproportionality in electoral systems affect policy outcomes. 
\cite{PolbornSnyder:2017} show that polarization may increase in legislatures when election outcomes become more uncertain. 
\cite{Prummer:2020} shows that a more fragmented media landscape leads to an increase in polarization when parties can micro-target voters during campaigns. 
\cite{DENTER:2021} shows that greater valence leads to policy moderation when there are complementarities between policy and valence. 
Similarly, \cite{BalartCasasTroumpounis:2022} study how modern information technology, which allows for more precise targeting of voters, may induce policy polarization. \cite{Yuksel:2022} shows that in fractionalized societies, policy polarization tends to be greater when voters must spend time learning about parties'  platforms. 
Taken together, this literature identifies several mechanisms through which institutional design, information frictions, and voter heterogeneity shape policy polarization. 
However, few papers isolate the role of ideological polarization itself when parties are both policy- and office-motivated. 
This paper aims to fill that gap.

The paper also contributes to the literature on the implications of differences in valence, pioneered by \cite{Stokes:1963}.  
\cite{AnsolabehereSnyder:2000}, \cite{Groseclose:2001}, and \cite{AragonesPalfrey:2002} study models of electoral competition with policy-motivated candidates and show that candidates with a valence advantage tend to choose more moderate positions than their disadvantaged counterparts. 
A paper closely related to the present one is \cite{BuisseretvanWeelden:2022}, who enrich the model of electoral competition between office-motivated candidates by introducing fixed candidate ideologies as a third characteristic. 
Their analysis reveals that if the weight of ideology in the voter's utility function is sufficiently high---i.e., if societies are polarized---then the standard result that the candidate with the valence advantage adopts the more moderate position may not hold. 
A similar result appears in \cite{Xefteris:2014}, who extends the classical model to three candidates and shows that when there is significant uncertainty about voter preferences, the candidate with the greatest valence may choose the most extreme policy position. 
\cite{ARAGONESXefteris:2017} consider a model in which voters differ in how they evaluate candidates' non-policy characteristics and demonstrate that policy polarization follows a U-shaped pattern with respect to the share of voters supporting the more popular candidate. 
The current paper is closely related to \cite{BuisseretvanWeelden:2022} but allows for policy motivation on the candidates' side and uncertainty about voter ideology as well as valence.
The analysis reveals that valence has no impact on the overall degree of policy polarization but determines which candidate locates closer to the political center, as in \cite{BuisseretvanWeelden:2022}. 
If societies are sufficiently polarized, the candidate or party with the valence advantage may take a \emph{less} extreme position.

Finally, the paper contributes to the literature studying electoral competition when positions on a subset of policy issues are fixed. 
The pioneering contribution in this field is due to \cite{KRASAPolborn:2010}, who study a model of electoral competition with office-motivated candidates. 
There are multiple binary policy issues, and candidates are exogenously committed to fixed positions on some issues while choosing positions on others. 
\cite{BuisseretvanWeelden:2020} analyze a variant of their model with two issues and two incumbent politicians who hold fixed and opposing positions on one of these dimensions but can freely choose policies in the other. 
Their key insight is that incumbents are particularly vulnerable to outsider entry when ideological polarization is high. 
\cite{hughes2025strategic} extends this framework in another direction to study legislative elections and shows that equilibrium policies in such settings are both more predictable and more representative of voter preferences than in single-district elections. 
The present paper differs in that, while parties occupy fixed ideological positions, they are free to choose policy in a second, continuous dimension. 
The focus is thus to highlight how ideological polarization shapes the polarization of policies.

\bigskip

The paper proceeds as follows. In Section~\ref{sec:model}, I introduce the model and define the equilibrium. Section~\ref{sec:results} establishes the main results regarding policy polarization. Section~\ref{sec:hetero} derives additional results on equilibrium party positioning as a function of valence advantages. Section~\ref{sec:conclusion} concludes. All formal proofs are relegated to the Appendix.

\section{Model}
\label{sec:model}
%The model is similar to the one proposed in \cite{BuisseretvanWeelden:2022}.
 Two parties, $j \in \{L, R\}$, compete in an election across two  dimensions. The first dimension captures \textit{party ideology}, along which the parties 
are assumed to be polarized. This dimension reflects the conventional cleavage between 
political parties, such as the divide between conservative and liberal positions. I denote positions in this dimension by $i\in\mathbb{R}$.
Party $L$ occupies $i_L=0$, while party $R$ has $i_R=1$. As in \cite{krasa2014social} or \cite{BuisseretvanWeelden:2022}, these ideological positions are fixed and cannot be altered, reflecting, for example, long-run partisan identities and reputational constraints that make repositioning along the core ideological dimension infeasible, at least in the short run.
The second dimension represents a policy issue such as redistribution or health policy, and I denote positions in this dimension by $p\in\mathbb{R}$. Parties have preferences over policies (to be defined precisely below), but  they are free to choose any policy $p_j\in\mathbb{R}$. Hence, like in  \cite{KRASAPolborn:2010,krasa2014social}, \cite{BuisseretvanWeelden:2020,BuisseretvanWeelden:2022}, or \cite{hughes2025strategic}, parties have a fixed position on one issue, and are flexible to choose any position on the other.

%Therefore, the  space of policies from which parties can choose is $\mathcal{S}=\{0,1\}\times[0,1]$.
%Figure \ref{fig:policyspace} shows the policy space and the respective strategy space for parties $L$ (blue) and $R$ (red).
%\begin{%figure}
%  \centering
%  \includegraphics[width=.6\textwidth]{fig_polspace.pdf}
%  \caption{Policy space $\{0,1\}\times [0,1]$ and parties' strategy spaces.}\label{fig:policyspace}
%\end{figure}

There is a single voter (she), who is  characterized by her vector of preferred policies, $(\hat{i}_V,\hat{p}_V)$. $\hat{p}_V$ is known by the parties and equals $\ot$. $\hat{i}_V\in\mathbb{R}$ is not precisely known by the parties, but there is common knowledge that $\hat{i}_V$ follows a normal distribution with mean $\mu_i=\ot$ and standard deviation $\sigma_i>0$.  

The voter's  utility from  an ideology-policy pair $(i,p)$ is
\[
u(i_j,p_j)=-w\left(\hat{i}_V-i_j\right)^2-\left(\hat{p}_V-p_j\right)^2.
\]
$w\geq 0$ is a parameter that measures the relative importance of ideology and policy in the voter's utility function. Following \cite{BuisseretvanWeelden:2022}, I interpret $w$ as the model's measure of ideological polarization between the two parties and in society at large. Since the ideological positions of the two parties are fixed at $0$ and $1$, a higher value of $w$ makes this ideological gap more consequential for voter utility and electoral choice. It therefore increases the extent to which voting decisions hinge on ideological alignment rather than on the flexible policy dimension. In this sense, a larger $w$ captures a more polarized environment in which ideology plays a greater role in voters' decisions.

Apart from ideology and policy, the voter cares about candidates' valence. $R$ has a valence advantage of $v\in\mathbb{R}$, which  follows a normal distribution and has mean $\mu_v=0$ and standard deviation $\sigma_v>\tilde{\sigma}_v\equiv\sqrt{\frac{32}{3125}}\approx 0.101$.\footnote{The assumption that $\sigma_v>\tilde{\sigma}_v$ is not necessary, but sufficient for the expected utility of each party to be single-peaked in the own policy choice.}
The voter  casts a ballot for party $L$ if
\[
u(i_L,p_L)>u(i_R,p_R)+v.
\]

As in \cite{Wittman:1983} or \cite{Calvert:1985}, parties are both  office motivated and policy motivated. In particular, if elected into office, a party receives office rents equal to $V>0$. Moreover, parties  receive both ideological and  policy utility, just like the voter.  Every party has as the most preferred ideology the own  ideological position, implying  that parties' ideological bliss points are $\hat{i}_L=0$ and $\hat{i}_R=1$. Moreover, parties' ideal policies are $\hat{p}_L=0$ and $\hat{p}_R=1$. Therefore, while parties may choose identical policies, their preferences differ. Parties' utility functions are as follows:
\[
\pi_j=\left\{
\begin{array}{rl}
V-w\left(\hat{i}_j-i_{j}\right)^2-\left(\hat{p}_j-p_j\right)^2=V-\left(\hat{p}_j-p_j\right)^2&\text{if party } j \text{ wins}\\
-w\left(\hat{i}_j-i_{-j}\right)^2-\left(\hat{p}_j-p_{-j}\right)^2=-w-\left(\hat{p}_j-p_{-j}\right)^2&\text{if party } -j \text{ wins}
\end{array}
\right.
\]

Parties choose policy platforms  to maximize their expected utility. The solution concept is Bayesian Nash equilibrium. Because the model is constructed in a way that no party has a competitive advantage, the resulting equilibrium will be symmetric, with $p_L^*=1-p_R^*$. This allows to derive results in a clear and parsimonious way. %

\bigskip

\paragraph{Discussion of Assumptions:}
Before solving the model, it is worthwhile to briefly pause and discuss some of the model's main assumptions.

The baseline model is set up so that no party has an advantage in the election \textit{in expectation}. This is, of course, not entirely realistic. However, it allows for clean results in a complex setting by analyzing the properties of the symmetric equilibrium. In this way, I can explain the mechanisms driving the main results in a clear and parsimonious manner. At the end of Section~\ref{sec:results}, I show that the main result on policy polarization is robust to small departures from symmetry.

Another assumption is that parties face uncertainty about the valence advantage and the voter's ideal ideology, while the voter's ideal policy platform is known. Allowing also for uncertainty about the voter's preferred policy would add an additional source of electoral uncertainty, but would not change the qualitative mechanism behind the main result. However, the resulting expressions become considerably more unwieldy, and thus, to improve exposition, I restrict attention to two sources of uncertainty.

Uncertainty about both valence and ideology is important for the non-monotonicity derived below. Proposition~\ref{prop:1} isolates the two effects: with only valence uncertainty, greater ideological polarization leads to more moderate policy platforms, whereas with only ideological uncertainty, it leads to greater policy divergence. The U-shaped relationship therefore results from the \textit{interaction} of these two sources of uncertainty.

\section{Ideological and Policy Polarization}
\label{sec:results}

To build intuition about the effect of increasing ideological polarization on platform choices, and to highlight the role of the two sources of uncertainty in the model, it is useful to first examine two polar cases: (i) $\sigma_v \rightarrow 0$, implying that the valence advantage is \textit{known} to be $v = 0$, and (ii) $\sigma_i \rightarrow 0$, implying that the voter's ideological position is \textit{known} to be $\hat{i}_V = \tfrac{1}{2}$.

If $\sigma_i \rightarrow 0$, then there is only valence uncertainty, while the voter's ideology is known. When $w$ is small, the voter cares mainly about $p$ and valence. Uncertainty regarding valence allows parties to move away from $p = \tfrac{1}{2}$ and closer to their own preferred policies. As $w$ increases, however, ideology becomes more important, which raises the cost of losing office, since defeat entails a larger ideological disutility. This makes parties behave as if they were more office-motivated, inducing them to adopt more centrist positions: platform polarization therefore \textit{decreases}.

Conversely, if $\sigma_v \rightarrow 0$,\footnote{Note that $\sigma_v\rightarrow 0$ contradicts the assumption that $\sigma_v> \tilde{\sigma}_v$. In the main model with both sources of uncertainty present, this assumption is useful to guarantee that the second-order conditions of the parties are satisfied. Here I dispense with this assumption. The reason is that the aim of the analysis of the two extreme cases is to show that the main results regarding policy polarization derived below exist only when both sources of uncertainty, valence and ideology, coexist.} then there is only ideological uncertainty, while the valence advantage is known to be zero. When $w$ is very small, both parties ignore ideology and choose the voter's known preferred policy, $\hat{p}_V = \tfrac{1}{2}$. As $w$ increases, uncertainty about ideology becomes increasingly relevant, allowing parties to move away from the electoral center. Hence, platform polarization \textit{increases}.

The first proposition formalizes these intuitions:

\begin{prop}
Consider the two polar cases in which there is only one source of uncertainty. Let $\Delta(w) = \left|p_R^*(w) - p_L^*(w)\right|$ denote policy polarization. Then:
\begin{itemize}
    \item If $\sigma_i \rightarrow 0$, policy polarization $\Delta(w)$ decreases in ideological polarization $w$ in any interior pure-strategy equilibrium.
    \item If $\sigma_v \rightarrow 0$, policy polarization $\Delta(w)$ increases in ideological polarization $w$ in any interior pure-strategy equilibrium.
\end{itemize}
\label{prop:1}
\end{prop}

Proposition~\ref{prop:1} suggests that the effect of increasing ideological polarization on policy polarization depends on which source of uncertainty dominates, and that the direction of the effect is not clear \textit{a priori}. Examining these extreme cases, however, helps develop intuition for the general case. Consider first the situation in which nobody cares about ideology, $w = 0$. Then, parties' chances of winning depend solely on valence $v$ and policy $p$. As $w$ increases from this point, ideological concerns begin to matter but are initially dominated by valence effects, because $w$ is ``small.'' Following Proposition~\ref{prop:1}, we may surmise that, for low levels of  $w$, parties choose increasingly moderate policy platforms as ideological polarization rises. However, as $w$ continues to grow, ideological motives become more important, and, in line with Proposition~\ref{prop:1}, eventually the comparative statics reverse: greater ideological polarization  leads parties to adopt more extreme platforms. Hence, when $w$ is large, ideological considerations dominate valence uncertainty, and greater ideological polarization induces parties to diverge further in policies. The resulting relationship between ideological polarization and policy polarization is therefore \textit{non-monotonic}. The next result formalizes this insight:

\begin{prop}
\label{prop:sym_EQ}
The game has a unique symmetric Nash equilibrium in pure strategies, in which $p_L^* = 1 - p_R^*$. Moreover, in this equilibrium, $p_L^*$ is non-monotonic and single-peaked in $w$. In particular, there exists $\tilde{w} > 0$ such that $p_L^*$ increases in $w$ for all $w \in [0, \tilde{w})$ and decreases in $w$ for all $w > \tilde{w}$.
\end{prop}

Parties' policy platforms are thus non-monotonic functions of ideological polarization $w$. Recall that our measure of policy polarization is $\Delta(w) = \left|p_R^*(w) - p_L^*(w)\right|=1 - 2p_L^*(w)$. It follows directly from Proposition~\ref{prop:sym_EQ} that policy polarization is also non-monotonic and U-shaped:

\begin{prop}
\label{prop:dDeltedw}
Let
\[
\Delta(w) = \left|p_R^*(w) - p_L^*(w)\right|.
\]
Policy polarization $\Delta(w)$ is a U-shaped function of ideological polarization $w$. Moreover,
\[
\Delta(0) = \frac{\sqrt{\sigma_v^2 + 4V^2 \phi(0)^2 + 4\sigma_v (V + 2)\phi(0)} - 2V\phi(0) - \sigma_v}{4\phi(0)} \in (0,1),
\]
and
\[
\lim_{w \rightarrow \infty} \Delta(w) = \frac{\sigma_i}{\sigma_i + \phi(0)} \in (0,1).
\]
\end{prop}

\begin{figure}
 \centering
 \includegraphics[width=.65\textwidth]{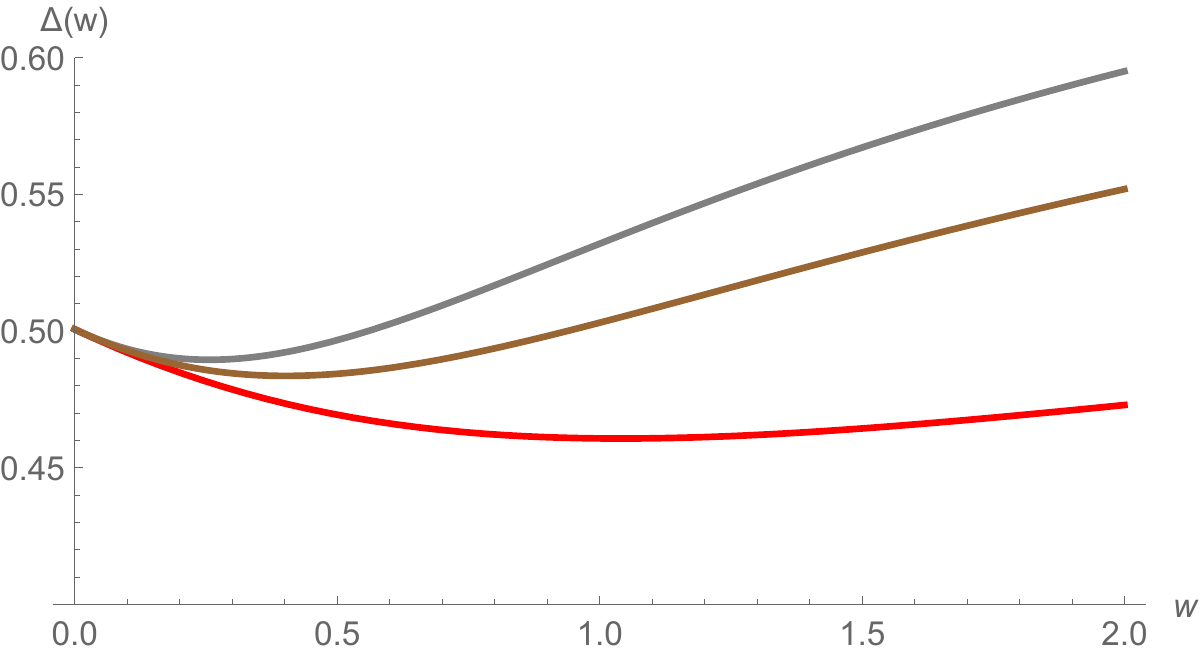}
 \caption{Platform polarization $\Delta(w)$ in the symmetric equilibrium as a function of $w \in [0,2]$ for varying $\sigma_i$.}
\label{fig:polarization}
\end{figure}

Policy polarization is  a U-shaped function of ideological polarization $w$, see also Figure~\ref{fig:polarization}. This mirrors the pattern shown in Figure~\ref{fig:side_by_side}, and the model thus offers a theoretical explanation for the evolution of policy polarization observed in the United States over the past three decades.

The analysis so far has focused on a symmetric environment in which neither party has an electoral advantage in expectation. This assumption allows for a particularly clean characterization of equilibrium policy polarization. The non-monotonicity, however, is not specific to exact symmetry. In particular, the U-shaped relationship continues to hold locally when parties differ slightly in their expected valence or ideological support. The next proposition establishes this robustness result.

\begin{prop}
\label{prop:asum_polarization}
 A small electoral advantage, whether due to valence or ideology, shifts both equilibrium policy platforms toward the advantaged party's preferred policy. Hence, the advantaged party adopts a more extreme position, while the disadvantaged party becomes more moderate. Moreover, the U-shaped relationship between ideological and policy polarization is robust to small asymmetries between the parties, and locally the degree of policy polarization is unaffected by these asymmetries.
\end{prop}

\section{Asymmetries and the Moderate Frontrunner}
\label{sec:hetero}
Proposition~\ref{prop:asum_polarization} shows that the main result on policy polarization is robust to small departures from symmetry. It also shows that while polarization is not affected by small changes, both parties change policies in the direction of the preferred policy of the now advantaged party.

Earlier studies have shown that differences in valence can lead to increased polarization, because the weaker party needs to differentiate itself from the stronger one and, hence, adopts a more extreme policy platform (see, for example, \cite{AnsolabehereSnyder:2000}, \cite{Groseclose:2001}, and \cite{AragonesPalfrey:2002}). \cite{DENTER:2021} shows that this result may even hold when parties are policy-motivated. \cite{Groseclose:2001} coined this the \textit{moderate frontrunner result}. However, \cite{BuisseretvanWeelden:2022} demonstrate that this prediction may not survive in a model with ideological differentiation of parties. In their setting, when ideological polarization is low, the moderate frontrunner result remains intact, and thus policy congruence and valence are positively correlated. However, if ideological polarization is sufficiently large, the prediction may reverse: if the high-valence party faces an ideological disadvantage, it may choose the more extreme policy position, implying that policy congruence and valence are negatively correlated.

The current setup allows for both valence and ideological advantages, which need not point in the same direction. A valence advantage may be offset by an ideological disadvantage, so that neither party has an electoral advantage overall. In that case, the equilibrium remains symmetric and both parties are equally distant from the electoral center:

\begin{prop}
The symmetric equilibrium described above exists for any $\mu_v = w(1 - 2\mu_i)$.
\label{prop:asym}
\end{prop}

\begin{figure}
   \centering
  \includegraphics[width=0.5\linewidth]{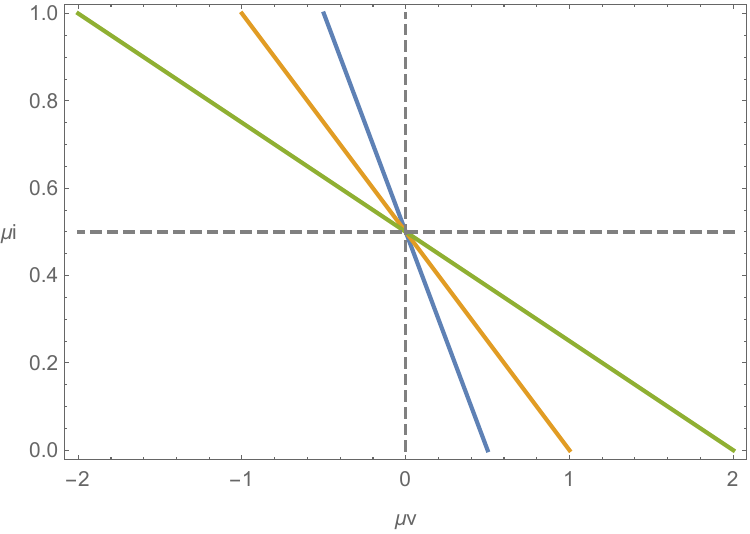}
 \caption{Combinations of $\mu_v$ and $\mu_i$ such that a symmetric equilibrium exists for $w \in \{\tfrac{1}{2}, 1, 2\}$.}
\label{fig:placeholder}
\end{figure}

In a symmetric equilibrium, both parties win the election with equal probability. The proposition therefore shows that expected valence advantages alone are insufficient to guarantee an electoral advantage and that, even with expected valence differences, a symmetric equilibrium may exist. For this to hold, an advantage in one dimension must be offset by a disadvantage in the other. The relative size of these compensating effects depends on the ideological polarization of society, $w$. When $w$ is large, even a small ideological advantage compensates for a large valence disadvantage, and vice versa. Figure~\ref{fig:placeholder} illustrates combinations of $\mu_v$ and $\mu_i$ for which this balance is achieved for varying $w$.

%A direct implication of Proposition~\ref{prop:asym} is that valence advantages alone do not reveal much about the degree of policy polarization. Different levels of valence asymmetries may lead to the same level of policy polarization for appropriately chosen levels of ideological polarization.

%Proposition~\ref{prop:asym} also implies that the moderate frontrunner result no longer holds if the frontrunner is defined as the party with a valence advantage. The next and final result of this paper examines what happens to policy platforms when we depart from the symmetric equilibrium. Will the party with the electoral advantage adopt the more moderate position, as suggested by the earlier literature, or should we expect the advantaged party to move closer to its own ideal policy?

Proposition~\ref{prop:asym} establishes that a symmetric equilibrium can persist even when one party enjoys an expected valence advantage, provided that this advantage is exactly offset by an ideological disadvantage. In other words, equilibrium symmetry hinges on the precise balance between valence and ideology. The next proposition examines what happens when this balance is disturbed, i.e., when the  condition $\mu_v = w(1 - 2\mu_i)$ no longer holds. In such asymmetric configurations, we can identify which party adopts the more moderate position and how this outcome depends on the degree of ideological polarization $w$.

\begin{prop}
Assume $\mu_v < 0$ and $\mu_i > \tfrac{1}{2}$. If $w > \hat{w} := \frac{\mu_v}{(1 - 2\mu_i)}$, then party $L$ adopts a more moderate policy platform than party $R$. Otherwise, if $w < \hat{w}$, then party $R$ adopts the more moderate platform than party $L$.
\label{prop:asym2}
\end{prop}

The proposition shows that the moderate frontrunner result does not hold in the current model. In fact, the opposite result emerges: a party with an electoral advantage chooses a more extreme platform. To see this, note that when $w = 0$, party $L$ has the electoral advantage. However, the proposition shows that in this case, party $R$ adopts the more moderate stance. This prediction thus contrasts with \cite{BuisseretvanWeelden:2022}, who find that under low polarization, the advantaged party remains moderate.

To understand this result, it is important to recall the key differences in modeling assumptions. In the current framework, the policy space is continuous, parties are partially policy-motivated, and the valence advantage is uncertain with infinite support, whereas the voter's preferred policy $\hat{p}$ is known. These assumptions imply that without policy motivation, parties would converge on the median of medians in the policy space to maximize their winning probabilities. Adding policy motivation causes both parties to move away from the electoral center. Moreover, any electoral advantage will be used to further \textit{both} objectives: winning office and promoting favorable policy outcomes. Consequently, electoral advantages translate into more extreme policy choices.

\begin{figure}
 \centering
\includegraphics[width=.95\textwidth]{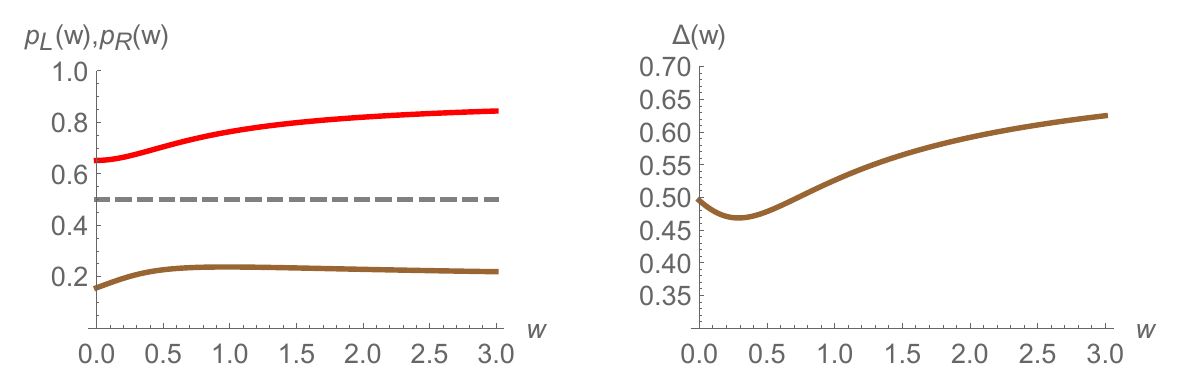}
  \caption{Platform choices as a function of $w$ when party $L$ has a valence advantage and party $R$ an ideological advantage (left panel). Platform polarization $\Delta(w)$ in the asymmetric equilibrium as a function of $w \in [0,3]$ (right panel).}
 \label{fig:asym}
\end{figure}

Generally, the party with the electoral advantage adopts the more extreme policy platform. The left panel of Figure~\ref{fig:asym} plots the equilibrium platform choices as a function of $w$ when $\mu_i = -\mu_v = V = \sigma_i = \sigma_v = 1$. The equilibrium is symmetric when $w = 1$. When $w$ is small, the ideological advantage of party $R$ is dominated by party $L$'s valence advantage, and $L$ chooses the more extreme platform. As $w$ increases, this reverses, and eventually $R$ becomes more extreme despite its valence disadvantage. Policy polarization is depicted in the right panel of Figure \ref{fig:asym} and remains U-shaped in $w$.

\section{Conclusion}
\label{sec:conclusion}
This paper develops a model of electoral competition in which two ideologically polarized parties are both office- and policy-motivated. The analysis shows that ideological and policy polarization need not move in parallel. When ideological polarization is low, an increase leads to policy moderation; when it is high, it amplifies policy divergence, yielding a U-shaped relationship between the two forms of polarization. This non-monotonicity is robust to small departures from symmetry. The model therefore provides one possible theoretical explanation for why increasing ideological polarization need not translate monotonically into more polarized policy platforms.

The analysis also speaks to the \textit{moderate frontrunner result} stated by \cite{Groseclose:2001}. A small electoral advantage shifts both equilibrium platforms toward the advantaged party's preferred policy, so that the advantaged party becomes more extreme while the disadvantaged party becomes more moderate. More generally, when ideological polarization is low, high-valence parties tend to adopt more extreme positions. When ideological polarization is sufficiently high, this pattern may reverse, as ideological advantages become more important for electoral competition. These findings highlight how ideological polarization changes both the degree of policy polarization and the way electoral advantages shape equilibrium party positions.

%In this paper I have analyzed how ideological polarization influences policy polarization in a game-theoretic model. Parties that are both office- and policy-motivated and polarized along an ideological dimension compete by offering policies to voters in order to secure election. The analysis advances our understanding of polarization by showing that ideological and policy polarization need not move in parallel. When ideological polarization is low, an increase reduces platform polarization, whereas when ideological polarization is high, further increases lead to greater policy polarization. Hence, the relationship between ideological and policy polarization is non-monotonic, mirroring the trends observed in the United States over the last 25 years (see Figure \ref{fig:side_by_side}).
%
%The paper also shows that valence levels and ideological polarization jointly determine which party adopts a more extreme and which a more centrist policy position, thereby confirming the main result of \cite{BuisseretvanWeelden:2022} in an extended framework.

%
% We show that a central prediction
%from the existing work on electoral competition—that higher valence candidates
%adopt positions that are more aligned with the electorate—may be reversed if polar-
%ization is large enough. We identify cases in which valence and policy congruence
%could be positively or negatively correlated. Given the increase in polarization in
%recent decades, this provides a framework to interpret empirical results on the link
%between policy moderation and electoral success.

\bigskip

\appendix

\section{Mathematical Appendix}
\subsection{Proof of Proposition \ref{prop:1}}
I prove the two parts of the proposition separately.
\paragraph{$\sigma_i\rightarrow 0$:} 
If the only uncertainty comes from valence, it is known that $\hat{i}_V=\ot$. It follows that $L$ wins the election iff
\[
-w\left(0-\ot\right)^2-\left(p_L-\ot\right)^2>-w\left(1-\ot\right)^2-\left(p_R-\ot\right)^2+v\Leftrightarrow v<p_L(1-p_L)-(1-p_R) p_R.
\]
Hence, the probability that $L$ wins the election is
\[
\text{Pr}=\Phi\left(\frac{p_L(1-p_L)-(1-p_R) p_R}{\sigma_v}\right),
\]
where $\Phi$ is the c.d.f of the standard Gaussian distribution. It follows that the parties' expected utilities are as follows:
\[
\begin{array}{rcl}
\mathbb{E}\left[\pi_L\right]&=&\text{Pr}\left(V-(0-p_L)^2\right)-\left(1-\text{Pr}\right)\left(w+(0-p_R)^2\right)\\
\mathbb{E}\left[\pi_R\right]&=&(1-\text{Pr})\left(V-(1-p_R)^2\right)-\text{Pr}\left(w+(1-p_L)^2\right)
\end{array}
\]
Define $\kappa:=\frac{p_L(1-p_L)-p_R(1-p_R)}{\sigma_v}$. Then:
\[
\dfrac{\partial \mathbb{E}\left[\pi_L\right]}{\partial p_L}=\dfrac{(1-2 p_L) \left(p_R^2-p_L^2+V+w\right) \phi\left(\kappa\right)}{\sigma_v}-2 p_L\Phi\left(\kappa\right)
\]
Because $\left.\frac{\partial \mathbb{E}\left[\pi_L\right]}{\partial p_L}\right|_{p_L=0}=\frac{\left(p_R^2+V+w\right) \phi\left(\frac{(p_R-1) p_R}{\sigma_v}\right)}{\sigma_v}>0$ and $\left.\frac{\partial \mathbb{E}\left[\pi_L\right]}{\partial p_L}\right|_{p_L=\frac{1}{2}}=-\Phi \left(-\frac{\left(\frac{1}{2}-p_R\right) \left(p_R-\frac{1}{2}\right)}{\sigma_v}\right)<0$, the optimal $p_L$ has to be interior, $p_L^*\in(0,\ot)$. Similarly, we can show that $p_R^*\in(\ot,1)$.

Because the game is completely symmetric, we focus on symmetric interior pure strategy equilibria.
Invoking symmetry, $p_R=1-p_L$, and using $\Phi(0)=\ot$, the FOC simplifies to
\begin{equation}
\label{FOC_SYM}
\left.\dfrac{\partial \mathbb{E}\left[\pi_L\right]}{\partial p_L}\right|_{p_R=1-p_L}=\dfrac{(1-2 p_L) \phi(0) (V+w+1-2 p_L)}{\sigma_v}- p_L
\end{equation}
It is straightforward to show that if $p_L$ solves $\left.\frac{\partial \mathbb{E}\left[\pi_L\right]}{\partial p_L}\right|_{p_R=1-p_L}=0$, then $p_R=1-p_L$ solves also $\left.\frac{\partial \mathbb{E}\left[\pi_R\right]}{\partial p_R}\right|_{p_L=1-p_R}=0$.  
Denote the equilibrium policy platform of party $L$ in a symmetric pure strategy equilibrium by $p_L^*$.

\begin{comment}
The second derivative of $\mathbb{E}[\pi_L]$ is
\[
\begin{array}{rcl}
\dfrac{\partial^2 \mathbb{E}[\pi_L]}{\partial p_L^2}&=&
-2 \Phi\left(\kappa\right)\\
&+&\frac{2 \sigma_v \left(p_L (5 p_L-2)-p_R^2-V-w\right) \phi\left(\kappa\right)-(1-2 {p_L})^2 \left({p_L}^2-{p_R}^2-V-w\right) \phi'\left(\kappa\right)}{{\sigma_v}^2}
\end{array}
\]
For large enough $\sigma_v$ this is strictly negative, and thus the SOC is satisfied globally. Hence, \eqref{FOC_SYM} characterizes the unique symmetric equilibrium. Moreover, evaluated at $p_L=1-p_R$,
\[
\left.\dfrac{\partial^2 \mathbb{E}[\pi_L]}{\partial p_L^2}\right|_{p_R=1-p_L}=-\frac{2 \phi(0) \left(1-4 p_L^2+V+w\right)}{\sqrt{\sigma_v^2+4 \sigma_i^2 w^2}}-1<0
\]
for any $p_L\in(0,\ot)$. Thus, locally the SOC is always satisfied when $\eqref{FOC_SYM}$ is equal to zero.

 IDIOT ME, WRONG PLACE FOR THAT PROOF
    
\end{comment}

To determine how $p_L$ changes with $w$, we use the implicit function theorem. Using $\Phi(0)=\ot$, we get
\[
\frac{\partial p_L^*}{\partial w}=\frac{(1-2 p_L^*) \phi(0)}{2 \phi(0) (V+w+2(1-2 p_L^*))+ \sigma_v}.
\]
Because $p_L^*\in(0,\ot)$, both the numerator and denominator are positive, and thus $\frac{\partial p_L^*}{\partial w}>0$ in any symmetric equilibrium. This proves the first part of the proposition.

\paragraph{$\sigma_v\rightarrow 0$:}
If the only uncertainty comes from ideology, it is known that $v=0$. It follows that $L$ wins the election iff
\[
-w\left(0-\hat{i}_V\right)^2-\left(p_L-\ot\right)^2>-w\left(1-\hat{i}_V\right)^2-\left(p_R-\ot\right)^2\Leftrightarrow \hat{i}_V<\frac{\left(p_L(1-p_L)-(1-p_R) p_R\right)+w}{2 w}.
\]
Hence, the probability that $L$ wins the election is
\[
\text{Pr}=\Phi \left(\frac{\frac{\left(p_L(1-p_L)-(1-p_R) p_R\right)+w}{2
   w}-\frac{1}{2}}{\sigma_i}\right),
\]
where as before $\Phi$ is the c.d.f of the standard Gaussian distribution. Parties' expected utilities are still as follows:
\[
\begin{array}{rcl}
\mathbb{E}\left[\pi_L\right]&=&\text{Pr}\left(V-(0-p_L)^2\right)-\left(1-\text{Pr}\right)\left(w+(0-p_R)^2\right)\\
\mathbb{E}\left[\pi_R\right]&=&(1-\text{Pr})\left(V-(1-p_R)^2\right)-\text{Pr}\left(w+(1-p_L)^2\right)
\end{array}
\]
Again focusing on symmetric equilibria, it suffices to calculate one FOC. Define $\kappa:=\frac{\frac{\left(p_L(1-p_L)-(1-p_R) p_R\right)+w}{2
   w}-\frac{1}{2}}{\sigma_i}$. Then:
\[
\dfrac{\partial \mathbb{E}\left[\pi_L\right]}{\partial p_L}=\frac{(1-2 p_L) \left(p_R^2-p_L^2+V+w\right) \phi
   \left(\kappa\right)}{2 \sigma_i w}-2 p_L \Phi \left(\kappa\right)
\]
Invoking symmetry, $p_R=1-p_L$, and using $\Phi(0)=\ot$, this simplifies to
\begin{equation}
\label{FOC_SYM2}
\left.\dfrac{\partial \mathbb{E}\left[\pi_L\right]}{\partial p_L}\right|_{p_R=1-p_L}=\frac{(1-2 p_L) \phi(0) (V+w+1-2 p_L)}{2 \sigma_i w}-p_L
\end{equation}
\begin{comment}
The second derivative is
\[
\begin{array}{rcl}
\dfrac{\partial^2 \mathbb{E}\left[\pi(m,p_i^m,p_2^m|\hat{i}_V^L,\hat{p}_V^L)\right]}{\partial p_L^2}&=&-2 \Phi (\kappa )\\
&+&\frac{(1-2 p_L)^2 \phi'(\kappa ) \left(p_R^2-p_L^2+V+w\right)-4
   \sigma_i w \phi(\kappa ) \left((2-5 p_L)
   p_L+p_R^2+V+w\right)}{4 \sigma_i^2 w^2}
\end{array}
\]
For large enough $\sigma_i$ this is strictly negative, and thus the SOC is satisfied. Hence, \eqref{FOC_SYM2} characterizes the unique symmetric equilibrium. Evaluated at $p_L=0$ this is $\frac{(V+w+1) \phi(0)}{2 \sigma_i w}>0$. When $p_L=\ot$, we get $-\ot$. Hence the pure strategy equilibrium is interior. Denote the equilibrium value by $p_L^*$.
    
\end{comment}

To determine how $p_L$ changes with $w$, we use again the implicit function theorem:
\[
\frac{\partial p_L^*}{\partial w}=-\frac{(1-2 p_L^*) (V+1-2 p_L^*) \phi(0)}{2 w \left(\phi(0) (V+w+2(1-2
   p_L^*))+\sigma_i w\right)}
\]
Because $p_L^*\in(0,\ot)$, both the numerator and denominator are positive, and thus $\frac{\partial p_L^*}{\partial w}<0$ in any symmetric equilibrium. This proves the second part of the proposition.
\qed
\subsection{Proof of Proposition \ref{prop:sym_EQ}}
 $L$ wins the election iff
\[
-w\left(0-\hat{i}_V\right)^2-\left(p_L-\ot\right)^2-v>-w\left(1-\hat{i}_V\right)^2-\left(p_R-\ot\right)^2\Leftrightarrow 2 w \, \hat{i}_V+v<p_L(1-p_L) - (1 - p_R) p_R + w
\]
$2 w \,\hat{i}_V+v$ is normally distributed with mean $2w\mu_i+\mu_v=w$ and standard deviation $\sqrt{\sigma_v^2+4w^2\sigma_i^2}$.
Hence, the probability that $L$ wins the election is
\[
\text{Pr}=\Phi \left(\frac{p_L(1-p_L)-(1-p_R) p_R}{\sqrt{\sigma_v^2+4w^2\sigma_i^2}}\right),
\]
where again  $\Phi$ is the c.d.f of the standard Gaussian distribution. Parties' expected utilities are:
\[
\begin{array}{rcl}
\mathbb{E}\left[\pi_L\right]&=&\text{Pr}\left(V-(0-p_L)^2\right)-\left(1-\text{Pr}\right)\left(w+(0-p_R)^2\right)\\
\mathbb{E}\left[\pi_R\right]&=&(1-\text{Pr})\left(V-(1-p_R)^2\right)-\text{Pr}\left(w+(1-p_L)^2\right)
\end{array}
\]
The first derivatives are
\[
\begin{array}{rcl}
\dfrac{\partial \mathbb{E}\left[\pi_L\right]}{\partial p_L}&=&\dfrac{(2 p_L-1) \phi(\kappa )
   \left(p_R^2-p_L^2+V+w\right)}{\sqrt{\sigma_v^2+4 \sigma_i^2 w^2}}-2 p_L \Phi (\kappa )\\
\dfrac{\partial \mathbb{E}\left[\pi_R\right]}{\partial p_R}&=&-\dfrac{(2 p_R-1) \phi(\kappa) ((p_L-2) p_L-(p_R-2)
  p_R+V+w)}{\sqrt{\sigma_v^2+4 \sigma_i^2 w^2}}+2\left[ (p_R-1) \Phi (\kappa )+1-
  p_R\right]
\end{array}
\]

Any equilibrium has to be interior, because
\[
\left.\frac{\partial E[\pi_L]}{\partial p_L}\right|_{p_L=0}
=
\frac{(p_R^2+V+w)\phi\left(
\frac{(p_R-1)p_R}
{\sqrt{\sigma_v^2+4\sigma_i^2w^2}}
\right)}
{\sqrt{\sigma_v^2+4\sigma_i^2w^2}}
>0,
\]
and
\[
\left.\frac{\partial E[\pi_L]}{\partial p_L}\right|_{p_L=\frac12}
=
-\Phi\left(
-\frac{
\left(\frac12-p_R\right)\left(p_R-\frac12\right)}
{\sqrt{\sigma_v^2+4\sigma_i^2w^2}}
\right)
<0.
\]
Similarly,
\[
\left.\frac{\partial E[\pi_R]}{\partial p_R}\right|_{p_R=1}
=
-\frac{
\left((p_L-2)p_L+V+w+1\right)
\phi\left(
-\frac{(p_L-1)p_L}
{\sqrt{\sigma_v^2+4\sigma_i^2w^2}}
\right)}
{\sqrt{\sigma_v^2+4\sigma_i^2w^2}}
<0,
\]
while
\[
\left.\frac{\partial E[\pi_R]}{\partial p_R}\right|_{p_R=\frac12}
=
1-\Phi\left(
-\frac{\left(p_L-\frac12\right)^2}
{\sqrt{\sigma_v^2+4\sigma_i^2w^2}}
\right)
>0.
\]

\begin{comment}

XXX

Any equilibrium has to be interior, because $\left.\frac{\partial \mathbb{E}\left[\pi_L\right]}{\partial p_L}\right|_{p_L=0}=\frac{\left(p_R^2+V+w\right) \phi\left(\frac{(p_R-1) p_R}{\sqrt{\sigma_v^2+4
   \sigma_i^2 w^2}}\right)}{\sqrt{\sigma_v^2+4 \sigma_i^2 w^2}}>0$, $\left.\frac{\partial \mathbb{E}\left[\pi_L\right]}{\partial p_L}\right|_{p_L=\ot}=-\Phi \left(-\frac{\left(\frac{1}{2}-p_R\right)
   \left(p_R-\frac{1}{2}\right)}{\sqrt{\sigma_v^2+4 \sigma_i^2 w^2}}\right)<0$, $\left.\frac{\partial \mathbb{E}\left[\pi_L\right]}{\partial p_R}\right|_{p_R=1}=-\frac{\left((p_L-2) p_L+V+w+1 \right)\phi\left(-\frac{(p_L-1) p_L}{\sqrt{\sigma_v^2+4 \sigma_i^2 w^2}}\right)}{\sqrt{\sigma_v^2+4 \sigma_i^2 w^2}}<0$, and $\left.\frac{\partial \mathbb{E}\left[\pi_L\right]}{\partial p_R}\right|_{p_R=1}=1-\Phi \left(-\frac{\left(p_L-\frac{1}{2}\right)^2}{\sqrt{\sigma_v^2+4 \sigma_i^2
   w^2}}\right)>0$.
\end{comment}

Now consider the second derivatives:
\[
\begin{array}{rcl}
\dfrac{\partial^2 \mathbb{E}\left[\pi_L\right]}{\partial p_L^2}&=&\frac{(1-2 p_L)^2 \phi'(\kappa )
   \left(p_R^2-p_L^2+V+w\right)}{\sigma_v^2+4 \sigma_i^2 w^2}-\frac{2 \phi(\kappa ) \left((2-5 p_L) p_L+p_R^2+V+w\right)}{\sqrt{\sigma_v^2+4
   \sigma_i^2 w^2}}-2 \Phi (\kappa )\\
\dfrac{\partial^2 \mathbb{E}\left[\pi_R\right]}{\partial p_R^2}&=&-\frac{(1-2 p_R)^2 \phi'(\kappa ) ((p_L-2) p_L-(p_R-2)
   p_R+V+w)}{\sigma_v^2+4 \sigma_i^2 w^2}-\frac{2 \phi(\kappa ) ((p_L-2)
   p_L+(8-5 p_R) p_R+V+w-2)}{\sqrt{\sigma_v^2+4 \sigma_i^2 w^2}}-2(1-\Phi (\kappa ))
\end{array}
\]
I show that $\frac{\partial^2 \mathbb{E}\left[\pi_L\right]}{\partial p_L^2}<0$ whenever the FOC is satisfied. This implies any local extremum is a maximum and that the party's expected utility function is single-peaked with a unique peak.  Moreover, since we already established that there cannot be a maximum at the  boundaries, it follows that the peak is in the  interior of $[0,\ot]$. We can prove the corresponding result for party $R$ using identical steps, and thus  here I only show the proof for party $L$.

From the FOC it follows that
\[
\Phi (\kappa )= -\frac{(2 p_L-1) \phi(\kappa ) \left(p_R^2-p_L^2+V+w\right)}{2
   p_L \sqrt{\sigma_v^2+4 \sigma_i^2 w^2}}
\]
Using this, the SOC, when the FOC is satisfied, becomes
\[
-\frac{(1-2 p_L)^2 \phi'(\kappa ) \left(p_L^2-p_R^2-V-w\right)}{\sigma_v^2+4
   \sigma_i^2 w^2}-\frac{\phi(\kappa ) \left((3-8 p_L)
   p_L^2+p_R^2+V+w\right)}{p_L \sqrt{\sigma_v^2+4 \sigma_i^2 w^2}}<0
\]
Next note that $\phi'(\kappa)/\phi(\kappa)=-\kappa$. Thus, the inequality becomes
\[
%\begin{array}{c}
\kappa\frac{(1-2 p_L)^2  \left(p_L^2-p_R^2-V-w\right)}{\sigma_v^2+4
   \sigma_i^2 w^2}-\frac{\left((3-8 p_L)
   p_L^2+p_R^2+V+w\right)}{p_L \sqrt{\sigma_v^2+4 \sigma_i^2 w^2}}<0
%\Leftrightarrow \kappa(1-2 p_L)^2  \left(p_L%^2-p_R^2-V-w\right)-\frac{\left((3-8 p_L)
%p_L^2+p_R^2+V+w\right)}{p_L}\sigma_s<0,
%\end{array}
\]
Using the definition of $\kappa$, we can further simplify:
\[
\begin{array}{c}
\dfrac{(1-2 p_L)^2 \left(p_L(1-p_L)-(1-p_R) p_R\right)
   \left(p_L^2-p_R^2-V-w\right)-\frac{\left(\sigma_v^2+4 \sigma_i^2
   w^2\right) \left((3-8 p_L) p_L^2+p_R^2+V+w\right)}{p_L}}{\left(\text{$\sigma
   $v}^2+4 \sigma_i^2 w^2\right)^{3/2}}<0\\
\Leftrightarrow 
(1-2 p_L)^2 \left(p_L(1-p_L)-(1-p_R) p_R\right)
   \left(p_L^2-p_R^2-V-w\right)-\frac{\left(\sigma_v^2+4 \sigma_i^2
   w^2\right) \left((3-8 p_L) p_L^2+p_R^2+V+w\right)}{p_L}<0
\end{array}
\]
Because $\frac{\left(\sigma_v^2+4 \sigma_i^2
   w^2\right) \left((3-8 p_L) p_L^2+p_R^2+V+w\right)}{p_L}>0$, for large enough $\sigma_n^2=\sigma_v^2+4w^2\sigma_i^2$, the inequality is satisfied, and the expression strictly decreases in $\sigma_n$. Hence, we make it as small as possible. Since $\sigma_n^2=\sigma_v^2+4w^2\sigma_i^2\geq \sigma_v^2$,
it is sufficient to evaluate the expression at $\sigma_n^2=\sigma_v^2$.
Then:

\[
\begin{array}{c}
(1-2 p_L)^2\left(p_L(1-p_L)-(1-p_R) p_R\right)
   \left(p_L^2-p_R^2-V-w\right)-\frac{ \left((3-8 p_L) p_L^2+p_R^2+V+w\right)}{p_L}\sigma_v^2<0\\
\Leftrightarrow(1-2 p_L)^2 \left(p_L(1-p_L)-(1-p_R) p_R\right)
   >\frac{ \left((3-8 p_L) p_L^2+p_R^2+V+w\right)}{p_L\left(p_L^2-p_R^2-V-w\right)}\sigma_v^2
\end{array}
\]
The RHS increases in both $V$ and $w$. Thus, the inequality is less likely to hold if both are large. Let $S=V+w$ and take the limit in which $S\rightarrow \infty$:
\[
\lim_{S\to\infty}
\frac{((3-8p_L)p_L^2+p_R^2+S)}
{p_L(p_L^2-p_R^2-S)}
\sigma_v^2
=
-\frac{\sigma_v^2}{p_L}.
\]
Thus, if
\[
(1-2 p_L)^2 \left(p_L(1-p_L)-(1-p_R) p_R\right)
   >-\frac{\sigma_v^2}{p_L},
\]
then party $L$'s expected utility is single-peaked.
The LHS increases in $p_R$, and thus we let $p_R=\ot$, which yields:
\[
\frac{1}{4} p_L(1 - 2 p_L)^4<\sigma_v^2.
\]
The LHS is  maximized when $p_L=\frac{1}{10}$, yielding 
\[
\sigma_v^2>\frac{32}{3125}\Leftrightarrow \sigma_v>\sqrt{\frac{32}{3125}}\equiv\tilde{\sigma}_v\approx 0.10119.
\]
Hence, if $\sigma_v\geq \tilde{\sigma}_v$, then each parties' expected utility is single-peaked in the own strategy.

Next we analyze the symmetric equilibrium. First, consider the FOCs again, evaluate when $p_L=1-p_R$. In this case, $\kappa=0$, and hence:
\begin{equation}
\label{eq:sym_FOC}
\begin{array}{rcl}
\left.\dfrac{\partial \mathbb{E}\left[\pi_L\right]}{\partial p_L}\right|_{p_R=1-p_L}&=&-\frac{(2 p_L-1) \phi(0) (-2 p_L+V+w+1)}{\sqrt{\sigma_v^2+4 \sigma_i^2
   w^2}}-p_L\\
\left.\dfrac{\partial \mathbb{E}\left[\pi_R\right]}{\partial p_R}\right|_{p_R=1-p_L}&=&\frac{(2 p_L-1) \phi(0) (-2 p_L+V+w+1)}{\sqrt{\sigma_v^2+4 \sigma_i^2
   w^2}}+p_L
\end{array}
\end{equation}
Clearly, $\left.\frac{\partial \mathbb{E}\left[\pi_L\right]}{\partial p_L}\right|_{p_R=1-p_L}=-\left.\frac{\partial \mathbb{E}\left[\pi_R\right]}{\partial p_R}\right|_{p_R=1-p_L}$, and hence if one is zero, so is the other. Because each party's expected utility is single-peaked and concave at the optimum, a symmetric equilibrium exists and is determined by these FOCs. 

To calculate how $w$ influences equilibrium platforms, we use the implicit function theorem:
\begin{equation}
\label{eq:dpdw}
\frac{\partial p_L^*}{\partial w}=\frac{(1-2 p_L) \phi(0) \left(4 \sigma_i^2 w (2
   p_L-V-1)+\sigma_v^2\right)}{\left(\sigma_v^2+4 \sigma_i^2 w^2\right) \left(2 \phi(0) (V+w+2(1-2 p_L))+\sqrt{\sigma_v^2+4
   \sigma_i^2 w^2}\right)}
\end{equation}
We have 
\[\left.\frac{\partial p_L^*}{\partial w}\right|_{w=0}=\frac{(1-2 p_L) \phi(0)}{2 (2(1-2p_L)+V) \phi(0)+\sigma_v}>0,
\]and thus when ideology is not important, $p_L^*$ increases in $w$. When $w$ is sufficiently large, the sign  changes. To see this, we reformulate \eqref{eq:dpdw}:
\[
\begin{array}{c}
\dfrac{(1-2 p_L) \phi(0) \left(4 \sigma_i^2 w (2
   p_L-V-1)+\sigma_v^2\right)}{\left(\sigma_v^2+4 \sigma_i^2 w^2\right) \left(2 \phi(0) (V+w+2(1-2 p_L))+\sqrt{\sigma_v^2+4
   \sigma_i^2 w^2}\right)}<0\\
\Leftrightarrow -\dfrac{4 \sigma_i^2 w (1-2
   p_L+V)-\sigma_v^2}{2 \phi(0) (V+w+2(1-2 p_L))+\sqrt{\sigma_v^2+4
   \sigma_i^2 w^2}}<0\\
\end{array}
\]
Because $p_L^*\in(0,\ot)$, this holds for large enough $w$. We can solve for $w$ such that this derivative is negative:
\begin{equation}
\label{COND:w}
w>\frac{\sigma_v^2}{4 \sigma_i^2 (1+V-2 p_L)}
\end{equation}
Note that the RHS is increasing in $p_L$. Thus, if \eqref{COND:w} is satisfied for some $w'$, it is satisfied for any $w>w'$. Together with the fact that $\left.\frac{\partial p_L^*}{\partial w}\right|_{w=0}>0$, this implies that $p_L^*$ is single-peaked in $w$, and that there exists $w'>0$ such that $p_L^*$ increases in $w$ for $w<w'$ and that it decreases in $w$ for $w>w'$.\qed

\subsection{Proof of Proposition \ref{prop:dDeltedw}}
The first part about polarization being U-shaped in $w$ follows directly from Proposition \ref{prop:sym_EQ}.

For the limit results, take the FOC in \eqref{eq:sym_FOC}. When $w=0$, the FOC becomes
\[
\frac{(2 p_L-1) (2 p_L-V-1)}{\sqrt{2 \pi } {\sigma_v}}-p_L=0
\]
and the result follows after simple algebra. Similarly, taking the limit as $w\rightarrow \infty$, we get
\[
-p_L \left(\frac{1}{\sqrt{2 \pi } {\sigma_i}}+1\right)+\frac{1}{2 \sqrt{2 \pi} {\sigma_i}}=0,
\]
and the limit result follows immediately.\qed

\subsection{Proof of Proposition \ref{prop:asum_polarization}}

Parties' expected utilities are as before. Define $\kappa:=
\frac{p_L(1-p_L)-p_R(1-p_R)+w(1-2\mu_i)-\mu_v}
{\sqrt{\sigma_v^2+4\sigma_i^2w^2}}$. The FOCs are
\[
\begin{array}{rcl}
\dfrac{\partial \mathbb{E}[\pi_L]}{\partial p_L}
&=&
-\dfrac{(2p_L-1)\phi(\kappa)(p_R^2-p_L^2+V+w)}
{\sqrt{\sigma_v^2+4\sigma_i^2w^2}}
-2p_L\Phi(\kappa),
\\[2mm]
\dfrac{\partial \mathbb{E}[\pi_R]}{\partial p_R}
&=&
-\dfrac{(2p_R-1)\phi(\kappa)((p_L-2)p_L-(p_R-2)p_R+V+w)}
{\sqrt{\sigma_v^2+4\sigma_i^2w^2}}
\\
&&
+2(p_R-1)\Phi(\kappa)-2p_R+2.
\end{array}
\]

Consider the Jacobian
\[
M=
\left(
\begin{array}{cc}
\dfrac{\partial^2 \mathbb{E}[\pi_L]}{\partial p_L^2}
&
\dfrac{\partial^2 \mathbb{E}[\pi_L]}{\partial p_L\partial p_R}
\\[2mm]
\dfrac{\partial^2 \mathbb{E}[\pi_R]}{\partial p_R\partial p_L}
&
\dfrac{\partial^2 \mathbb{E}[\pi_R]}{\partial p_R^2}
\end{array}
\right).
\]
At the symmetric equilibrium, $\mu_v=0$, $\mu_i=\frac{1}{2}$, and
$p_R=1-p_L$. Using the equilibrium FOC, the elements of $M$ simplify to
\[
\begin{array}{rcl}
\dfrac{\partial^2 \mathbb{E}[\pi_L]}{\partial p_L^2}
=
\dfrac{\partial^2 \mathbb{E}[\pi_R]}{\partial p_R^2}
&=&
\frac{1}{2p_L-1}
-
\frac{4p_L^2}{V+w+1-2p_L},
\\[2mm]
\dfrac{\partial^2 \mathbb{E}[\pi_L]}{\partial p_L\partial p_R}
=
\dfrac{\partial^2 \mathbb{E}[\pi_R]}{\partial p_R\partial p_L}
&=&
-\frac{2p_L(2p_L-1)}{V+w+1-2p_L}.
\end{array}
\]
Hence, $|M|\neq 0$. The implicit function theorem therefore implies that,
in a neighborhood of the symmetric equilibrium, $p_L^*$ and $p_R^*$ are
continuously differentiable functions of $w$, $\mu_v$, and $\mu_i$.
Consequently, equilibrium policy polarization
\[
\Delta(w,\mu_v,\mu_i)=p_R^*(w,\mu_v,\mu_i)-p_L^*(w,\mu_v,\mu_i)
\]
is continuous in all three arguments. Hence, for sufficiently small
asymmetries, polarization is arbitrarily close to polarization in the
symmetric case, and the interior minimum of $\Delta(w)$ is preserved.
%This proves the first part of the proposition.

Next, consider the effect of a small asymmetry on equilibrium policies.
For $x\in\{\mu_v,\mu_i\}$, define
\[
M_L^x=
\left(
\begin{array}{cc}
-\dfrac{\partial^2 \mathbb{E}[\pi_L]}{\partial p_L\partial x}
&
\dfrac{\partial^2 \mathbb{E}[\pi_L]}{\partial p_L\partial p_R}
\\[2mm]
-\dfrac{\partial^2 \mathbb{E}[\pi_R]}{\partial p_R\partial x}
&
\dfrac{\partial^2 \mathbb{E}[\pi_R]}{\partial p_R^2}
\end{array}
\right)
\]
and
\[
M_R^x=
\left(
\begin{array}{cc}
\dfrac{\partial^2 \mathbb{E}[\pi_L]}{\partial p_L^2}
&
-\dfrac{\partial^2 \mathbb{E}[\pi_L]}{\partial p_L\partial x}
\\[2mm]
\dfrac{\partial^2 \mathbb{E}[\pi_R]}{\partial p_R\partial p_L}
&
-\dfrac{\partial^2 \mathbb{E}[\pi_R]}{\partial p_R\partial x}
\end{array}
\right).
\]
By Cramer's rule,
\[
\frac{\partial p_L^*}{\partial x}
=
\frac{|M_L^x|}{|M|}
\qquad\text{and}\qquad
\frac{\partial p_R^*}{\partial x}
=
\frac{|M_R^x|}{|M|}.
\]

At the symmetric equilibrium, $\kappa=0$ and $\phi'(0)=0$. Hence,
\[
\frac{\partial^2 \mathbb{E}[\pi_L]}{\partial p_L\partial \mu_v}
=
\frac{\partial^2 \mathbb{E}[\pi_R]}{\partial p_R\partial \mu_v}
=
\frac{2p_L\phi(0)}
{\sqrt{\sigma_v^2+4\sigma_i^2w^2}},
\]
and
\[
\frac{\partial^2 \mathbb{E}[\pi_L]}{\partial p_L\partial \mu_i}
=
\frac{\partial^2 \mathbb{E}[\pi_R]}{\partial p_R\partial \mu_i}
=
\frac{4wp_L\phi(0)}
{\sqrt{\sigma_v^2+4\sigma_i^2w^2}}.
\]
Thus, for either $x=\mu_v$ or $x=\mu_i$,
\[
\frac{\partial p_L^*}{\partial x}
=
\frac{\partial p_R^*}{\partial x}>0.
\]
An increase in $\mu_v$ or $\mu_i$ increases party $R$'s electoral advantage
and shifts both equilibrium platforms toward party $R$'s preferred policy.
Hence, party $R$ becomes more extreme while party $L$ becomes more moderate.
Analogously, an electoral advantage of party $L$ shifts both platforms
toward party $L$'s preferred policy.

Finally,
\[
\frac{\partial \Delta}{\partial x}
=
\frac{\partial p_R^*}{\partial x}
-
\frac{\partial p_L^*}{\partial x}
=0,
\]
for $x\in\{\mu_v,\mu_i\}$. Thus, locally, small asymmetries affect the
location of equilibrium policy platforms, but not the degree of policy
polarization. This proves the proposition.
\qed

\subsection{Proof of Proposition \ref{prop:asym}}
Follows directly from the fact that the FOCs are the same as before when $\mu_v=w(1-2\mu_i)$ and $p_R=1-p_L$, and the same single-peakedness argument applies.\qed

\subsection{Proof of Proposition \ref{prop:asym2}}
Parties' expected utilities look just as before. Define $\kappa:=\frac{p_L(1-p_L)-p_R(1-p_R)+w(1-2\mu_i)-\mu_v}{\sqrt{{\sigma_v}^2+4 \sigma_i^2 w^2}}$. The FOCs are as follows:
\[
\begin{array}{rcl}
\dfrac{\partial \mathbb{E}[\pi_L]}{\partial p_L}&=&-\frac{(2 p_L-1) \phi(\kappa ) \left(p_R^2-p_L^2+V+w\right)}{\sqrt{{\sigma_v}^2+4 \sigma_i^2 w^2}}-2 p_L
   \Phi (\kappa )\\
\dfrac{\partial \mathbb{E}[\pi_R]}{\partial p_R}&=&-\frac{(2 p_R-1) \phi(\kappa ) ((p_L-2) p_L-(p_R-2) p_R+V+w)}{\sqrt{{\sigma_v}^2+4 \sigma_i^2 w^2}}+2
   (p_R-1) \Phi (\kappa )-2 p_R+2
\end{array}
\]
When $\mu_v=w(1-2\mu_i)$, there is a symmetric equilibrium with $p_R=1-p_L$, implying $\kappa=0$. The FOC in determining the equilibrium policy platforms in this equilibrium is:
\[
\left.\dfrac{\partial \mathbb{E}[\pi_L]}{\partial p_L}\right|_{p_R=1-p_L\wedge \kappa=0}=-\frac{(2 p_L-1) \phi(0) (V+w+1-2 p_L)}{\sqrt{\sigma_v^2+4 \sigma_i^2 w^2}}-p_L\\
%\left.\dfrac{\partial \mathbb{E}u_R(p_L,p_R)}{\partial p_L}\right|_{p_R=1-p_L\wedge \kappa=0}&=&p_L-\frac{(2 pL-1) \phi(0) (2 p_L-V+w-1)}{\sqrt{{\sigma_v}^2+4 {\sigma_i}^2 w^2}}
\]
From here we can see that 
\begin{equation}
\phi(0)=\frac{p_L \sqrt{{\sigma_v}^2+4 {\sigma_i}^2 w^2}}{(2 p_L-1) (2 p_L-V-w-1)}.    
\label{eq:simplify}
\end{equation}
We will use this later.

Next, we calculate the comparative statics with respect to $w$ at the symmetric equilibrium. For this we need to calculate the following:
\[
M=\left(\begin{array}{cc}
\dfrac{\partial^2 \mathbb{E}[\pi_L]}{\partial p_L^2}&\dfrac{\partial^2 \mathbb{E}[\pi_L]}{\partial p_L\partial p_R}\\
\dfrac{\partial^2 \mathbb{E}[\pi_R]}{\partial p_L\partial p_R}&\dfrac{\partial^2 \mathbb{E}[\pi_R]}{\partial p_R^2}\\
\end{array}\right)
\]
as well as
\[
\begin{array}{ccc}
M_L=\left(\begin{array}{cc}
-\dfrac{\partial^2 \mathbb{E}[\pi_L]}{\partial p_L\partial w}&\dfrac{\partial^2 \mathbb{E}[\pi_L]}{\partial p_L\partial p_R}\\
-\dfrac{\partial^2 \mathbb{E}[\pi_R]}{\partial p_R\partial w}&\dfrac{\partial^2 \mathbb{E}[\pi_R]}{\partial p_R^2}\\
\end{array}\right)
&
\text{ and }
&
M_R=\left(\begin{array}{cc}
\dfrac{\partial^2 \mathbb{E}[\pi_L]}{\partial p_L^2}&-\dfrac{\partial^2 \mathbb{E}[\pi_L]}{\partial p_L\partial w}\\
\dfrac{\partial^2 \mathbb{E}[\pi_R]}{\partial p_R\partial p_L}&-\dfrac{\partial^2 \mathbb{E}[\pi_R]}{\partial p_R\partial w}\\
\end{array}\right),
\end{array}
\]
each evaluated at the symmetric equilibrium with $p_L^*=1-p_R^*$. Using \eqref{eq:simplify}, the different derivatives are
\[
\begin{array}{rcl}
\dfrac{\partial^2 \mathbb{E}[\pi_L]}{\partial p_L^2}&=&\frac{1}{2 p_L-1}-\frac{4 p_L^2}{V+w+1-2 p_L}\\
\dfrac{\partial^2 \mathbb{E}[\pi_L]}{\partial p_L\partial p_R}&=&-\frac{2 p_L (2 p_L-1)}{V+w+1-2 p_L}\\
\dfrac{\partial^2 \mathbb{E}[\pi_R]}{\partial p_L\partial p_R}&=&-\frac{2 p_L (2 p_L-1)}{V+w+1-2 p_L}\\
\dfrac{\partial^2 \mathbb{E}[\pi_R]}{\partial p_R^2}&=&\frac{1}{2 p_L-1}-\frac{4 p_L^2}{V+w+1-2 p_L}
\end{array}
\]
Moreover,
\[
\begin{array}{rcl}
\dfrac{\partial^2 \mathbb{E}[\pi_L]}{\partial p_L\partial w}&=&\frac{p_L (4 (\mu_i-1) p_L+1)}{(2 p_L-1) (V+w+1-2 p_L)}+\frac{4 p_L \sigma_i^2 w}{\sigma_v^2+4 \sigma_i^2 w^2}\\
\dfrac{\partial^2 \mathbb{E}[\pi_R]}{\partial p_R\partial w}&=&\frac{p_L (4 \mu_i p_L-1)}{(2 p_L-1) (-2 p_L+V+w+1)}-\frac{4 p_L \sigma_i^2 w}{\sigma_v^2+4
   \sigma_i^2 w^2}
\end{array}
\]
It holds  that $\frac{\partial p_L^*}{\partial w}+\frac{\partial p_R^*}{\partial w}=\frac{\left|M_L\right|+\left|M_R\right|}{\left|M\right|}$. Using the derivatives, we have
\[
\frac{\partial p_L^*}{\partial w}+\frac{\partial p_R^*}{\partial w}=\frac{4 (2 \mu_i-1) p_L^2}{1-16 p_L^3+12 p_L^2-4 p_L+V+w}.
\]
The denominator is strictly positive, and  thus the sign of the expression depends on how $\mu_i$ compares to $\ot$. By assumption, $\mu_i>\ot$ and thus $\frac{\partial p_L^*}{\partial w}+\frac{\partial p_R^*}{\partial w}>0$, implying that $p_L^*+p_R^*>1\Leftrightarrow p_R^*-\ot>\ot-p_L^*$. Thus, $R$ takes a more extreme position than $L$ despite $L$ having the valence advantage in a neighborhood of the symmetric equilibrium.

\medskip

Next, I show that a symmetric equilibrium can exist only under the condition stated in Proposition \ref{prop:asym}. Evaluated when $p_R=1-p_L$, the FOCs become
\[
\begin{array}{rcl}
\left.\dfrac{\partial \mathbb{E}[\pi_L]}{\partial p_L}\right|_{p_R=1-p_L}&=&-\frac{(2 p_L-1) (1-2 p_L+V+w) \phi\left(\frac{w-\mu_v-2 {\mu_i}
   w}{\sqrt{\sigma_v^2+4 \sigma_i^2 w^2}}\right)}{\sqrt{\sigma_v^2+4 \sigma_i^2 w^2}}-2 p_L \Phi \left(\frac{w-\mu_v-2 \mu_i w}{\sqrt{\sigma_v^2+4 \sigma_i^2 w^2}}\right)\\
&&
   \\
\left.\dfrac{\partial \mathbb{E}[\pi_R]}{\partial p_R}\right|_{p_R=1-p_L}&=&\frac{(2 p_L-1) (1-2 p_L+V+w) \phi\left(\frac{w-\mu_v-2 {\mu_i}
   w}{\sqrt{\sigma_v^2+4 \sigma_i^2 w^2}}\right)}{\sqrt{\sigma_v^2+4 \sigma_i^2 w^2}}-2 p_L \Phi \left(\frac{w-\mu_v-2 \mu_i w}{\sqrt{\sigma_v^2+4 \sigma_i^2 w^2}}\right)+2 p_L
\end{array}
\]
At a symmetric equilibrium, both of these FOCs need to be equal to zero. Hence, it has to hold that 
\[
\left.\dfrac{\partial \mathbb{E}[\pi_L]}{\partial p_L}\right|_{p_R=1-p_L}+\left.\dfrac{\partial \mathbb{E}[\pi_R]}{\partial p_R}\right|_{p_R=1-p_L}=2 p_L-4 p_L \Phi \left(\frac{w-\mu_v-2 {\mu_i}
   w}{\sqrt{{\sigma_v}^2+4 {\sigma_i}^2 w^2}}\right)=0.
\]
This is only possible if either $p_L=0$, which cannot be the case because
\[
\left.\dfrac{\partial \mathbb{E}[\pi_L]}{\partial p_L}\right|_{p_R=1-p_L\wedge p_L=0\wedge \kappa=0}=\frac{\phi(0) (V+w+1)}{\sqrt{\sigma_v^2+4 \sigma_i^2 w^2}}>0,
%\left.\dfrac{\partial \mathbb{E}u_R(p_L,p_R)}{\partial p_L}\right|_{p_R=1-p_L\wedge \kappa=0}&=&p_L-\frac{(2 pL-1) \phi(0) (2 p_L-V+w-1)}{\sqrt{{\sigma_v}^2+4 {\sigma_i}^2 w^2}}
\]
or if 
\[
\frac{w-{\mu_v}-2 {\mu_i} w}{\sqrt{{\sigma_v}^2+4 {\sigma_i}^2
   w^2}}=0\Leftrightarrow \mu_v=w(1-2\mu_i),
\]
which is the condition stated in Proposition \ref{prop:asym}. It follows that there can only be a symmetric equilibrium if $\mu_v=w(1-2\mu_i)\Leftrightarrow w=\hat{w}=\frac{\mu_v}{1-2\mu_i}$. 

Starting at $w=\hat{w}$ and increasing $w$, we know that $L$ chooses the more moderate platform than $R$. This must remain true for any $w>\hat{w}$. Because by continuity, if $R$  were to eventually adopt  the more moderate platform than $L$, there would need to be a $w$ such that there is again a symmetric equilibrium. But this is impossible, because $w>\hat{w}$. 
Similarly, starting at $w=\hat{w}$ and decreasing $w$, we know that $R$ chooses the more moderate platform than $L$. This must remain true for any $w<\hat{w}$. Because by continuity, if $L$  were to eventually adopt  the more moderate platform than $R$, there would need to be a $w$ such that there is again a symmetric equilibrium. But this also is impossible, because $w<\hat{w}$. This proves the proposition.
\qed

\newpage

\bibliography{mybib}{}
\bibliographystyle{chicago}

\end{document}